\documentclass[conference]{IEEEtran}

\usepackage{cite}
\usepackage{graphicx}
\usepackage{graphicx}
\usepackage{amsmath}
\usepackage{float}
\usepackage[spaces,hyphens]{url}
\usepackage[colorlinks,allcolors=blue]{hyperref} 
\usepackage[dvipsnames]{xcolor}
\usepackage[]{algorithm2e}
\usepackage{algpseudocode}
\usepackage{listings}
\usepackage{subcaption}

\ifCLASSINFOpdf
\else
\fi

\begin{document}
%

\title{Fuzzy Inference System for Test Case Prioritization in Software Testing}


\author{\IEEEauthorblockN{Aron Karatayev , {Anna Ogorodova },
Pakizar Shamoi\IEEEauthorrefmark{1}}
\IEEEauthorblockA{School of Information Technology and Engineering \\
Kazakh-British Technical University\\
Almaty, Kazakhstan\\
Email: 
\IEEEauthorrefmark{1}p.shamoi@kbtu.kz,
}
}


%


\maketitle



%
\IEEEpeerreviewmaketitle



\begin{abstract}

In the realm of software development, testing is crucial for ensuring software quality and adherence to requirements. However, it can be time-consuming and resource-intensive, especially when dealing with large and complex software systems. Test case prioritization (TCP) is a vital strategy to enhance testing efficiency by identifying the most critical test cases for early execution. This paper introduces a novel fuzzy logic-based approach to automate TCP, using fuzzy linguistic variables and expert-derived fuzzy rules to establish a link between test case characteristics and their prioritization. Our methodology utilizes two fuzzy variables—failure rate and execution time—alongside two crisp parameters: Prerequisite Test Case and Recently Updated Flag. Our findings demonstrate the proposed system's capacity to rank test cases effectively through experimental validation on a real-world software system. The results affirm the practical applicability of our approach in optimizing the TCP and reducing the resource intensity of software testing.

\end{abstract}

\section{Introduction}




Software testing has become increasingly important in recent years, with regression testing holding significant importance. Its main purpose is to ensure that changes, such as bug fixes or new features, do not introduce any new defects into previously validated code.

However, the regression testing process can often be labor-intensive and costly due to the number of test cases and the need to re-run them frequently. Given these constraints, the industry has turned to prioritization techniques to strategically order test cases, focusing initially on the most critical ones.

Developing an effective prioritization methodology, however, is not a trivial task. It needs to balance different factors,  such as test execution time, the likelihood of finding defects, and the historical data of test failures. Additionally, many consistent metrics and evaluation criteria for prioritization can be applicable for certain cases but unsuitable for others. Developing an efficient prioritization strategy that is suitable for different projects is a challenging task.  Therefore, a more flexible and convenient test case prioritization (TCP) system is needed to provide an efficient solution.


Currently, TCP systems underscore the significance of this process in software testing and highlight the potential of fuzzy logic, static parameters, and other algorithms like GA to enhance its effectiveness. However, more research is required to address the identified limitations and to explore potential synergies between mentioned approaches. 

The current paper provides a methodology for automating the prioritization process by using fuzzy sets and logic.
Fuzzy logic is applicable to this problem due to its ability to handle ambiguity and uncertainty, integrate expert knowledge, and combine multiple criteria into a resultant assessment.
The main paper's contribution is capturing experts' knowledge in constructing the fuzzy inference system, specifically fuzzy sets and fuzzy rules. Additionally, analysis and evaluation were conducted on a real manually collected dataset from real industrial projects. 





The paper is structured as follows. This Introduction is Section I. A review of the research on TCP is presented in Section II. Methodology, including fuzzy logic and data collection, are covered in Section III. Section IV presents the study results. Finally, concluding remarks and future works are provided in Section V. 

\section{Related Works}
TCP is a critical aspect of software testing, ensuring that tests are executed in a sequence that maximizes the early detection of defects and thus optimizes the use of testing resources and time. Various research has explored different methodologies of TCP. 

Traditional TCP techniques include basic prioritization strategies involving test adequacy criteria, such as code coverage, to represent different test case behaviours\cite{new15}. In one study, a technique for prioritization in regression testing was proposed \cite{Singh2005}, providing basic concepts of the TCP focused on testing changed code parts as soon as possible.  Similarly, the other study introduced an optimized regression testing approach that highlights other possible metrics for prioritization, like the Average Percentage of Fault Detected (APFD), and discusses the need for optimization test case usage \cite{Ansari2016}. Model transformations in Model-Driven Engineering (MDE) require efficient regression testing due to the high cost of executing many test cases. Techniques focusing on test case rule coverage information have been proposed to prioritize test cases during regression testing of model transformations\cite{new18}. Another study introduced an ACO-based approach to optimize the order of test cases in regression testing based on the number of defects detected, execution time, and defect severity \cite{new17}.  The other study proposed the concept of dynamic test case reordering in which the order of test cases is adjusted during regression testing in response to events occurring during testing and the most recent test set \cite{new13}.



Collective intelligence algorithms, particularly the Artificial Fish School Algorithm (AFSA) \cite{new19}, have been used to solve TCP problems. These algorithms use average test point coverage percentage and adequate execution time to optimize the test case selection process, showing significant improvements in single-target and multi-target TCP problems. Techniques such as Cuckoo-ACO, Epistemic Particle Swarm Optimisation and Leaping Frog Algorithm have been used to address the challenging nature of MOTCP (multiple optimization TCP)\cite{new19} by focusing on aspects such as efficient execution time and biological epistasis. Nature-inspired algorithms, such as the hybrid cuckoo search algorithm \cite{new20}, combine the advantages of different approaches to balance the trade-off between exploration and utilization. 


Next, defect-based approaches prioritize test cases based on their ability to detect defects \cite{new9}. Mutation testing, a defect-based technique, introduces defects into the original program to create mutants. Test cases are prioritized based on their ability to "kill" these mutants. 

In continuous integration (CI) environments, where software changes are often integrated, TCP plays a vital role \cite{new21}. Most approaches in CI are history-based, relying on the history of failures and test execution. Highly configurable systems present unique challenges for TCP due to the sheer number of configurations\cite{new22}. Approaches in this area often focus on functional and non-functional targets, with results showing that multi-target prioritization typically leads to faster defect detection compared to mono-target prioritization. 

The other study presents an approach that prioritizes test cases based on historical failure data, test execution time, and domain-specific heuristics \cite{new4}. Another study introduced a new automated TCP technique specifically for web applications \cite{new11}, focusing on identifying and prioritizing test cases related to database changes. The approach uses a dependency graph of functionality and prioritizes test cases based on their relevance to changed database elements, aiming for early defect detection.

Other methods include the use of Markov chains  \cite{new16},   modified genetic algorithm \cite{Yadav2017},  \cite{new14}, abstract test case prioritization (ATCP) techniques \cite{new10}, a new metric based on the probability of error propagation to rank test cases in order of their probabilistic error-finding ability \cite{new3}.

The application of fuzzy logic in TCP systems is another area of interest due to its potential to handle uncertainty and vagueness \cite{cw2}. For example, in \cite{Al-Refai2017}, authors proposed a fuzzy logic-based model for prioritization considering changes in the UML diagram and another based on code changes and compared both.  Other studies used software agents and fuzzy logic to prioritize test cases \cite{new5}, showing the potential benefits of this approach.

The field of test case prioritization has seen a variety of approaches, from traditional strategies focusing on code coverage to more complex and sophisticated techniques using collective intelligence algorithms, hybrid natural algorithms, and error-based methods. The choice of technique often depends on the specific needs and constraints of the test environment, including factors such as the size and nature of the software, the presence of multiple targets, and the specific characteristics of the test infrastructure. 

Integrating fuzzy logic and other advanced computational techniques paves the way for a more robust and reliable software testing framework.

\section{Methodology}
The main idea of our research is to develop a convenient TCP system and implement it to increase the time efficiency of the software testing process. The system aims to enhance the effectiveness of software testing by scheduling the execution of the test cases in an optimal order. This section outlines the specific methods used within the proposed methodology.

\subsection{Fuzzy Sets and Logic Theory}

This subsection provides an overview of the fundamental terms from fuzzy theory that we use in our research. 
\subsubsection{Membership Functions and Fuzzy Sets}

A fuzzy set is a class of objects with a continuum of grades of membership ranging between 0 and 1. A fuzzy set \(A\) in a universe of discourse \(X\) is characterized by a membership function \( \mu_A(x) \) which associates with each element \(x\) in \(X\) a real number in the interval \([0, 1]\), representing the degree of membership of \(x\) in \(A\) \cite{Zadeh1965}, \cite{Zadeh1975}. Mathematically, it is expressed as:
\[ \mu_A : X \rightarrow [0, 1] \]
where $ \mu_A(x) $ = degree of membership of  $x$ in $A$.

\subsubsection{Fuzzy Operations}
The $\alpha $-cut (Alpha cut) is a crisp set that includes all the members 
of the given fuzzy subset f whose values are not less than $\alpha $ for 0< 
$\alpha  \quad  \le $ 1 \cite{Zadeh1965}, \cite{Zadeh1975}:
\[
f_{\alpha}  = \{x:\mu _{f} (x) \ge \alpha \}
\]
To connect $\alpha $-cuts and 
set operations (A and B are fuzzy sets):
\[
(A \cup B)_{\alpha}  = A_{\alpha}  \cup B_{\alpha}  ,
\quad
(A \cap B)_{\alpha}  = A_{\alpha}  \cap B_{\alpha}  
\]
\subsubsection{Fuzzy Rules}
Fuzzy rules are a key component of fuzzy logic, a form of multi-valued logic derived from fuzzy set theory to deal with approximate reasoning. Fuzzy rules express logical relationships between inputs and outputs of a system in a way that mimics human reasoning. These rules are typically expressed in the format:
\text{IF } x \text{ is } A \text{ THEN } y \text{ is } B, where 
$x$ is the input variable, $A$ and $B$ are fuzzy sets defined on discourse's input and output universes, respectively \cite{Zadeh1965}, \cite{Zadeh1975}. For example, \textit{IF failure rate is high THEN priority is high}. In a fuzzy inference system, the rule base consists of a collection of IF-THEN rules.


\subsection{Dataset collection}

Our testing methodology necessitates a diverse and comprehensive pool of test cases. During the extensive exploration process, over 150 test cases from three distinct systems were undertaken. 
As a result of this work, we focused on only one system. Within this system, we further refined our dataset to include 48 well-structured and representative test cases. This selection was based on factors such as test case relevance, diversity, and complexity. The chosen test cases were then updated and formatted accordingly to ensure their compatibility with the TCP system. Table \ref{dataset} presents a sample of the test cases dataset used in the experiment.

\begin{table}[]
 \centering
    \caption{Test cases Dataset (e-commerce system).} 
    \label{dataset}
\begin{tabular}{|r|l|r|r|r|c|}
\hline
\multicolumn{1}{|l|}{\textbf{\#}} & \textbf{Test Case} & \multicolumn{1}{l|}{\textbf{\begin{tabular}[c]{@{}l@{}}Exec. \\ time\end{tabular}}} & \multicolumn{1}{l|}{\textbf{\begin{tabular}[c]{@{}l@{}}Failure \\ rate\end{tabular}}} & \multicolumn{1}{l|}{\textbf{Prerequisite}} & \multicolumn{1}{l|}{\textbf{\begin{tabular}[c]{@{}l@{}}Recently\\ updated\end{tabular}}} \\ \hline
 1        & Log in email       & 10                                                                                     & 20                                                                                    & 3                                          & false                                                                                    \\ \hline
2                                & Log in phone       & 15                                                                                     & 28                                                                                    & 3                                          & true                                                                                     \\ \hline
3                                & Register           & 45                                                                                     & 40                                                                                    & \multicolumn{1}{l|}{}                      & false                                                                                    \\ \hline
4                                & Open profile       & 5                                                                                      & 5                                                                                     & 1                                          & false                                                                                    \\ \hline
5                                & Log Out            & 5                                                                                      & 2                                                                                     & 1 / 2                                      & false                                                                                    \\ \hline
6                                & Retrieve main page  & 5                                                                                      & 60                                                                                    & \multicolumn{1}{l|}{}                      & false                                                                                    \\ \hline
7                                & Open Category      & 10                                                                                     & 20                                                                                    & \multicolumn{1}{l|}{}                      & false                                                                                    \\ \hline
8                                & Search item        & 15                                                                                     & 17                                                                                    & \multicolumn{1}{l|}{}                      & false                                                                                    \\ \hline
9                                & Retrieve item page  & 10                                                                                     & 15                                                                                    & 8                                          & true                                                                                     \\ \hline
10                               & Add to cart        & 15                                                                                     & 5                                                                                     & 9                                          & false                                                                                    \\ \hline
11                               & Open Cart          & 5                                                                                      & 11                                                                                    & \multicolumn{1}{l|}{}                      & false                                                                                    \\ \hline
12                               & Cart Update        & 20                                                                                     & 11                                                                                    & 10, 11                                     & false                                                                                    \\ \hline
13                               & Remove from Cart   & 10                                                                                     & 5                                                                                     & 10                                         & false                                                                                    \\ \hline
14                               & Add to favorites   & 20                                                                                     & 1                                                                                     & \multicolumn{1}{l|}{}                      & false                                                                                    \\ \hline
15                               & Retrieve favorites  & 15                                                                                     & 18                                                                                    & \multicolumn{1}{l|}{}                      & false                                                                                    \\ \hline
16                               & Compare items      & 40                                                                                     & 38                                                                                    & \multicolumn{1}{l|}{}                      & false                                                                                    \\ \hline
17                               & Checkout           & 80                                                                                     & 46                                                                                    & 10                                         & false                                                                                    \\ \hline
18                               & Track order        & 30                                                                                     & 14                                                                                    & 17                                         & false                                                                                    \\ \hline
19                               & Cancel Order       & 50                                                                                     & 6                                                                                     & 17                                         & false                                                                                    \\ \hline
20                               & Refund order       & 90                                                                                     & 37                                                                                    & 17                                         & false                                                                                    \\ \hline
\end{tabular}
\end{table}


\subsection{Proposed Approach}

\begin{figure*}[tb]
\centerline{\includegraphics[width=\textwidth]{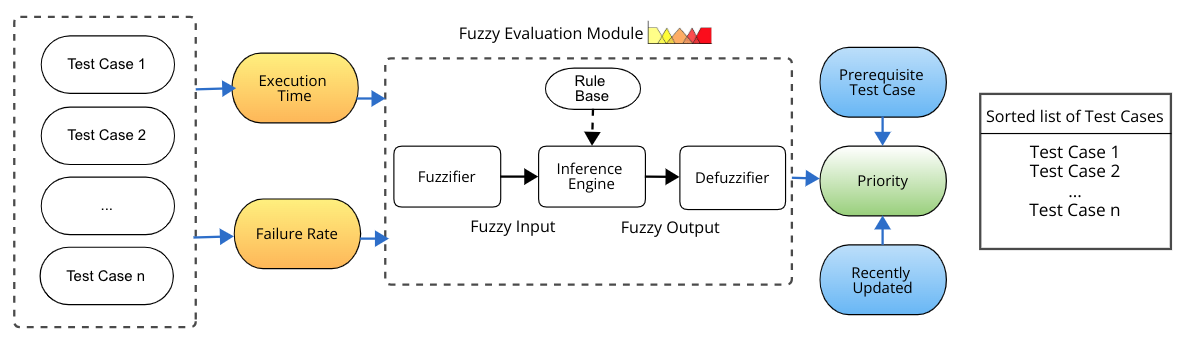}}
\caption{General Workflow of Test Cases Prioritization System.}
\label{main}
\end{figure*}

Fig. \ref{main} presents the proposed fuzzy logic-based approach. Our methodology uses two fuzzy variables—\textit{Failure rate} and \textit{Execution time}—and two crisp parameters: \textit{Prerequisite test case} and \textit{Recently updated} flag. 
%

\begin{figure*}[tb]
  \begin{subfigure}{0.33\textwidth}
    \includegraphics[width=\linewidth]{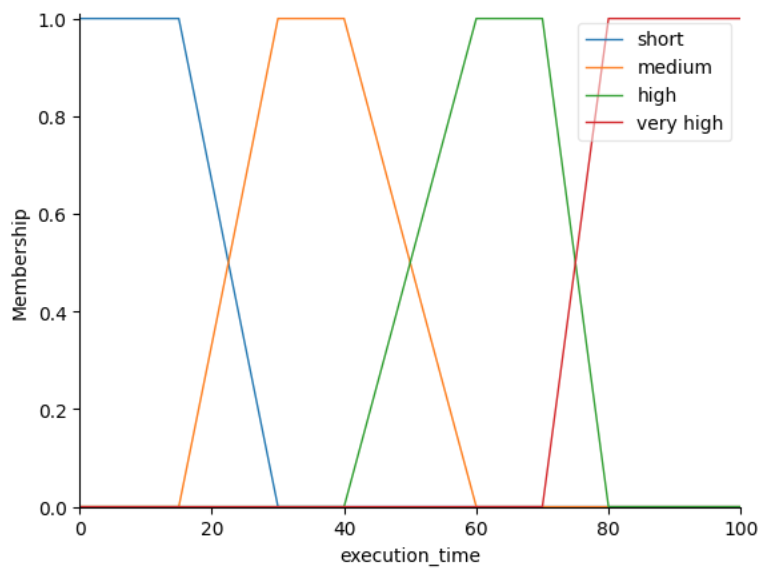}
  \end{subfigure}%
  \begin{subfigure}{0.33\textwidth}
    \includegraphics[width=\linewidth]{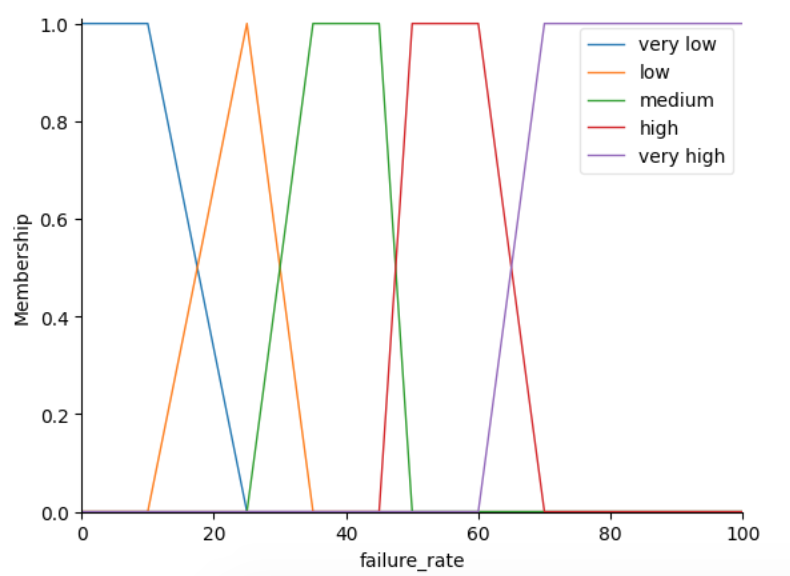}
  \end{subfigure}
  \begin{subfigure}{0.33\textwidth}
    \includegraphics[width=\linewidth]{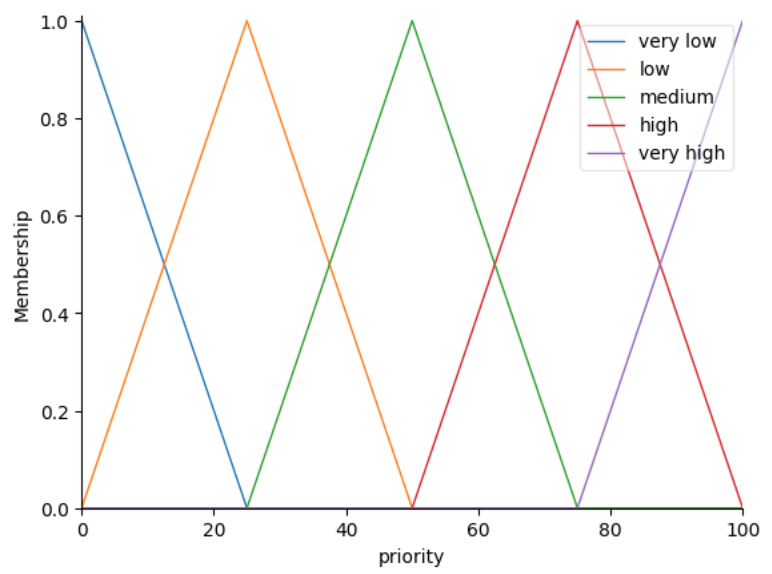}
  \end{subfigure}
 \caption{Input fuzzy sets for \textit{Execution Time}, \textit{Failure rate} and Output fuzzy sets for \textit{Priority }} 
\label{fig:fuzzy_sets}
\end{figure*}

Fuzzy logic provides a simple framework for dealing with imprecise and ambiguous information. We can assign fuzzy membership values to each test case, indicating their relative importance, by considering various prerequisite parameters like execution time and previous failure history. The test cases are then ranked according to these membership values, with higher values denoting higher priority.

 We represent \textit{Execution Time} as an ordered list of terms of the fuzzy variable \textit{X = "Execution Time"} using primary linguistic terms \textit{L = $\{$Short, Medium, High, Very High$\}$}, specifying its level. Similarly, the fuzzy variable \textit{Y = "Failure Rate"} is represented by terms \textit{L = $\{$Very low, Low, Medium, High, Very High$\}$}. Fig.~\ref{fig:fuzzy_sets} shows fuzzy input and output variables - the fuzzy sets for the \textit{Execution Time} and \textit{Failure Rate} variables and the output variable \textit{Priority}, represented by \textit{L = $\{$Very low, Low, Medium, High, Very High$\}$}. We use triangular and trapezoidal membership functions representing the linguistic labels expressing the fuzzy variable.  

We initially defined fuzzy sets subjectively. One of the advancements we made later was the definition of the fuzzy partition based on the findings of a survey we conducted with QA experts. Three experts participated in this process. Building membership functions based on experts' opinions enabled us to bridge the gap that existed between the system and experts' knowledge. 

So, the fuzzy partition was built using the Direct Rating method from \cite{cunn}. Various methods can be used to question the user to construct fuzzy sets. The Direct Rating method was chosen because it is simple, efficient, and requires fewer converging iterations. In Direct Rating, the user must choose a value from a list of possible linguistic values corresponding to a specific linguistic variable. In our experiment, experts (QA engineers) were asked to categorize different values into one of several categories corresponding to terms. The collected personal judgments were then used to define fuzzy sets. The tailored functions were obtained by limiting the shape of fuzzy sets to triangular or trapezoidal membership functions, as shown in Fig. \ref{fig:fuzzy_sets}. 



Expert-derived fuzzy rules are used to establish a relationship between test case characteristics and prioritization.   Table \ref{rules} presents the fuzzy rules. As can be seen, each rule considers the fuzzy sets of\textit{ Execution Time} and \textit{Failure Rate} and outputs a \textit{Priority} level for the test case. The fuzzy inference system processes these rules to determine the priority of each test case.

 \begin{table}[tb]
    \centering
    \caption{Fuzzy rules composed by the experts for the inference system.} 
    \begin{tabular}{|l|l|l|l|}
    \hline
      \textbf{  Rules} & \textbf{Execution time }& \textbf{Failure rate} &\textbf{ Prority} \\ \hline
        Rule 1 & High & Very Low & Very Low  \\ \hline
        Rule 2 & Very High & Very Low & Very Low  \\ \hline
        Rule 3 & Short & Very Low & Low  \\ \hline
        Rule 4 & Medium & Very Low & Low  \\ \hline
        Rule 5 & High & Low & Low  \\ \hline
        Rule 6 & Very High & Low & Low  \\ \hline
        Rule 7 & High & Medium & Low  \\ \hline
        Rule 8 & Very High & Medium & Low  \\ \hline
        Rule 9 & Short & Low & Medium  \\ \hline
        Rule 10 & Medium & Low & Medium  \\ \hline
        Rule 11 & Short & Medium & Medium  \\ \hline
        Rule 12 & Medium & Medium & Medium  \\ \hline
        Rule 13 & High & High & Medium  \\ \hline
        Rule 14 & Very High & High & Medium  \\ \hline
        Rule 15 & Short & High & High  \\ \hline
        Rule 16 & Medium & High & High  \\ \hline
        Rule 17 & High & Very High & High  \\ \hline
        Rule 18 & Very High & Very High & High  \\ \hline
        Rule 19 & Short & Very High & Very High  \\ \hline
        Rule 20 & Medium & Very High & Very High  \\ \hline
    \end{tabular}
     \label{rules}
\end{table}






\begin{table}[]
\caption{Parameters used for evaluation. 
}

\label{tablePars}
\begin{tabular}{|l|l|l|p{2cm}|}
\hline
\textbf{Parameter }             & \textbf{Type }   & \textbf{Fuzzy/Crisp} & \textbf{Description} \\ \hline
Execution Time         & int     & Fuzzy       &    The amount of time it takes for a specific task    \\ \hline
Failure Rate           & double  & Fuzzy       &       How often the test case fails under normal conditions    \\ \hline
Prerequisite Test Case & int     & Crisp       &      A test case that needs to be run first      \\ \hline
Recently Updated       & boolean & Crisp       &      A test case with  recently modified functions     \\ \hline

\end{tabular}

\end{table}
  In addition to the fuzzy logic components, the system also considers two crips variables: \textit{Prerequisite Test Case} and \textit{Recently Updated Flag} (see Table \ref{tablePars}). Crisp set specifies value as either 0 or 1, just like in classical binary logic.

The \textit{Prerequisite Test Case} parameter checks whether a test case has a prerequisite test case that needs to be run first. Prerequisite parameters played a crucial role in optimizing the test run flow. We recognized that some test cases had to be run multiple times to test different flows. Therefore, we incorporated the prerequisite parameter to establish a soft hierarchy among the test cases, allowing us to minimize the number of test runs. This approach also enabled us to ensure that all prerequisite test cases were run before the ones that depended on them.

The \textit{Recently Updated Flag} marks test cases that test recently modified functions. This parameter is included based on the assumption that recently changed functions are more likely to contain faults. Test cases marked with this flag are given a higher priority in the system; if the associated function was recently updated, the priority level was raised to the next level.


The workflow of the proposed methodology is as follows:
\begin{enumerate}

    \item \textbf{Input.} An array of test cases is received, each defined by a set of parameters. These include the failure rate, execution time, a prerequisites array, and a recently updated flag. 
    \item \textbf{Prioritization.} Upon receiving the input, the methodology begins the prioritization process based on fuzzy logic. Each test case's execution time and failure rate are fuzzified and passed to the fuzzy inference system to eventually compute its priority. If a test case has been marked as recently updated, it gets promoted to a higher priority level.
    \item \textbf{Outputs.} Finally, after all test cases have been evaluated and prioritized, we output a sequence of test cases. This output represents the optimized order in which the test cases should be executed.
\end{enumerate}

\section{Results}

\subsection{Performance Evaluation}
\begin{figure*}[tb]
  \begin{subfigure}{0.32\textwidth}
    \includegraphics[width=\linewidth]{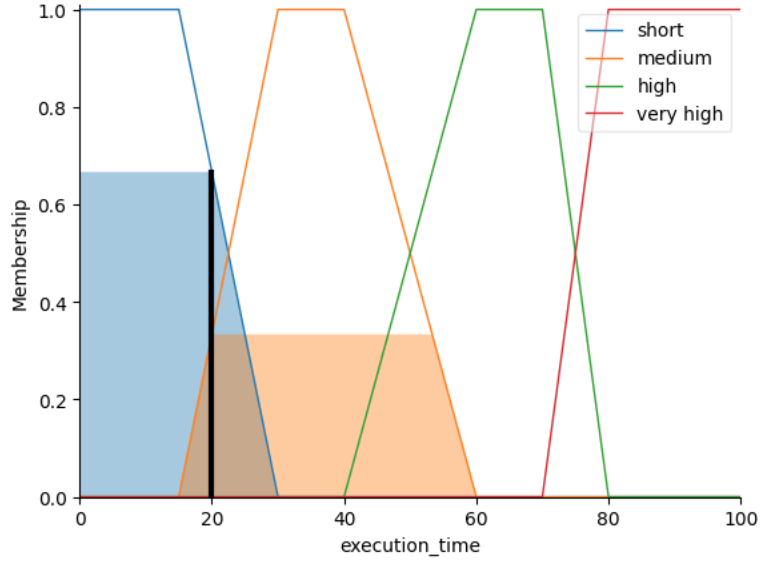}
    \caption{Applying input 20 sec on \textit{Execution Time} fuzzy set} \label{fig:1a}
  \end{subfigure}%
  \hspace*{\fill}   
  \begin{subfigure}{0.32\textwidth}
    \includegraphics[width=\linewidth]{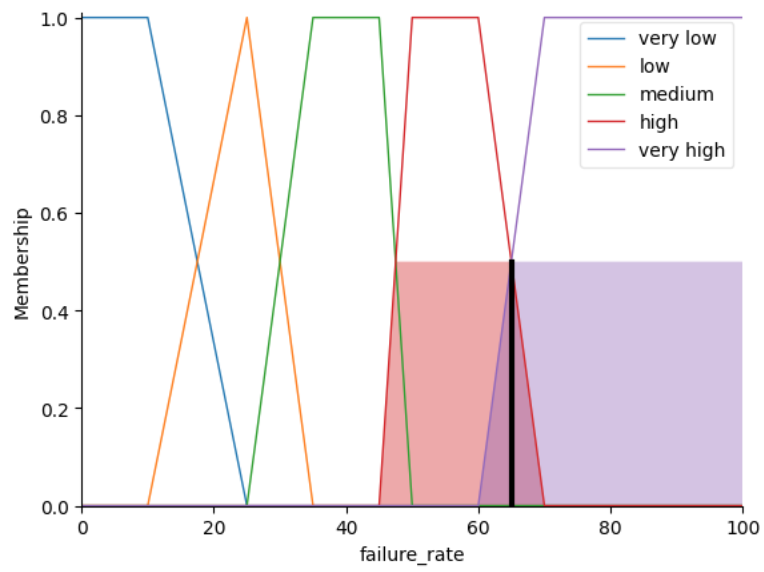}
    \caption{Applying input 65\% on \textit{Failure Rate} fuzzy set} \label{fig:1b}
  \end{subfigure}%
  \hspace*{\fill}   
  \begin{subfigure}{0.32\textwidth}
    \includegraphics[width=\linewidth]{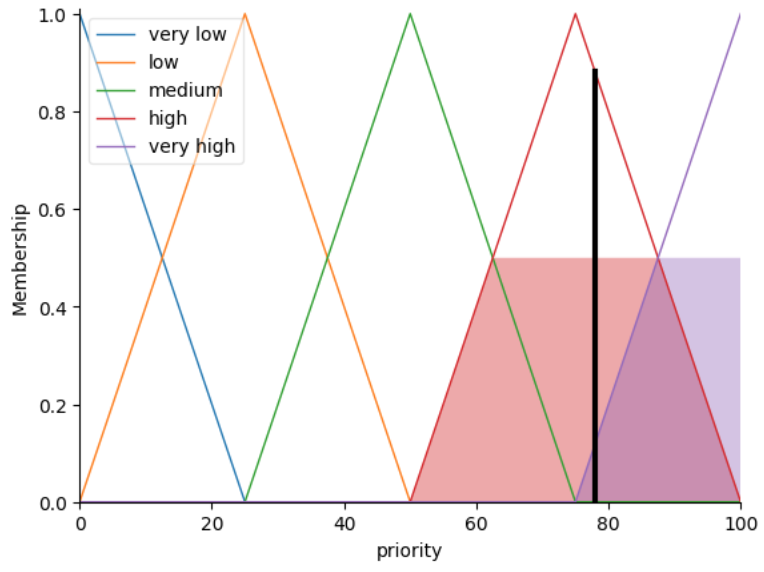}
    \caption{Aggregated Membership and Result, 77.98\%} 
  \end{subfigure}
\caption{Simulation Results.} \label{fig_ex}
\end{figure*}

A comparative study was performed to evaluate the effectiveness of the proposed fuzzy logic-based TCP system using three lists of test cases: an unsorted list (the way tests were executed before), a list sorted by a QA engineer, and the list formed based on the proposed methodology. The experiment results are presented in Table \ref{restable}. 

Three key metrics were used to assess the ordering in each list:
\begin{enumerate}
  \item The number of executed test cases. This metric is critical as certain flows necessitate running the same test cases multiple times. Efficient prerequisite mechanics should optimize this number.
  \item The total time (in seconds) spent running the tests.
  \item The number of errors that were identified during testing.
\end{enumerate}

\begin{table}[tb]
\caption{Evaluation Results}
\begin{tabular}{|l|r|r|r|}
\hline
\textbf{}                                     & \multicolumn{1}{l|}{\textbf{Unsorted}} & \multicolumn{1}{l|}{\textbf{QA expert}} & \multicolumn{1}{l|}{\textbf{Our methodology}} \\ \hline
 Executed tests amount & 32                                     & 25                                          & 27                                            \\ \hline 
Spent time (total)                                    & 725 sec                                & 605 sec                                     & 630 sec                                       \\ \hline

Failures found                                & 5                                      & 5                                           & 5                                             \\ \hline
\end{tabular}
\label{restable}
\end{table}

The most important test case was "Retrieve main page," which scored 75, while the one with the lowest priority was "Remove from cart," which scored 20.89. The fuzzy methodology provided satisfactory results compared to manual sorting done by QA specialists closely involved in the project. The proposed methodology significantly reduces the time and effort required by QA engineers.

Despite the room for improvement, the system demonstrated promising results compared to the unsorted list and list formed by an expert. This improvement represents a positive step forward, justifying further investigation and refinement when applying fuzzy logic to TCP.


\subsection{Example}




To simulate the fuzzy system, we must first specify the inputs and then use the defuzzification approach. Consider the following example: the \textit{Execution Time} and \textit{Failure Rate} are 20\% and 65\%, respectively (see Fig. \ref{fig_ex}). Fuzzy aggregation then aggregates the output membership functions using the maximum operator. Next, we must defuzzify to acquire a priority, which we accomplish using the centroid approach. Aggregation using fuzzy criteria yielded an overall \textit{Priority} of 77.98\%. Fig. \ref{fig_ex} depicts the visualized simulation results.

\section{Conclusion}

This paper presents a novel fuzzy logic-based approach for automating test case prioritization. It uses fuzzy linguistic variables and expert-derived fuzzy rules to build a link between test case attributes and prioritization. Our methodology combines two fuzzy variables (Failure Rate and Execution Time) with two crisp parameters (Prerequisite Test Case and Recently Updated Flag). This combination enabled us to benefit from both the fuzzification and fixed hierarchy approaches. 

The methodology is highly comparable with manual testing, with the intention of enhancing the efficiency of junior QA specialists in particular. Our experiments on a real-world software system show that the suggested system can efficiently rank test cases.


The key advantage of the methodology compared to similar studies is that real experts participated in knowledge base generation.

The proposed approach has certain limitations. Specifically, two of the main limitations are sensitivity to membership function design and weak generalization across projects. The fuzzy sets and rules specified for one project may not be directly relevant to the other. So, customization is required for each new project or software system, including the need to collect deep domain knowledge and expert insights. This process can be complex and time-consuming. The fuzzy logic-based prioritization system can adapt to different contexts only with the help of expert knowledge.

As for future work, we aim to improve the methodology by including additional prioritization parameters. We plan to integrate this system with the existing automated testing frameworks. This will help streamline the collection of essential parameters such as failure rate and execution time. We plan to extract these parameters directly from test run logs provided by the framework, which will streamline data acquisition and refine our prioritization process.




\bibliography{export}

\end{document}